\documentstyle[prb,aps,epsf,floats,goodfloat]{revtex}

\newcommand{\bc}{\begin{center}}
\newcommand{\ec}{\end{center}}
\newcommand{\be}{\begin{equation}}
\newcommand{\ee}{\end{equation}}
\newcommand{\beqn}{\begin{eqnarray}}
\newcommand{\eeqn}{\end{eqnarray}}
\begin{document}
\draft

\twocolumn[\hsize\textwidth\columnwidth\hsize\csname@twocolumnfalse%
\endcsname

\title{
Monte Carlo Study of the Critical Behavior of Random Bond Potts Models 
}

\author{T. Olson and A. P. Young}
\address{Department of Physics, University of California, Santa Cruz, 
CA 95064}

\date{\today}

\maketitle

\begin{abstract}
    We present results of Monte Carlo simulations of random bond Potts models
    in two dimensions, for different numbers of Potts states, $q$.  We
    introduce a simple scheme which yields continuous self-dual distributions
    of the interactions.  As expected, we find multifractal behavior of the
    correlation functions at the critical point and obtain estimates of the
    exponent $\eta_n$ for several moments, $n$, of the correlation functions,
    including {\em typical} ($n\to 0$), {\em average} ($n=1$) and others.  In
    addition, for $q=8$, we find that there is only a single correlation
    length exponent $\nu$ describing the correlation length away from
    criticality.  This is numerically very close to the pure Ising value, $\nu
    =1$. 
\end{abstract}

\vskip 0.3 truein
]

\section{Introduction}
In the theoretical study of phase transitions in disordered systems, it is
neither practical nor useful to give a description of a single sample with a
particular realization of the disorder. Rather, a detailed theory would yield a
{\em probability distribution} of the possible results for, say, the free
energy, obtained by sampling over {\em all} possible realizations of
the disorder. However, one generally argues that important quantities, such as
the free energy, are self averaging, that is to say independent of the
particular realization of the disorder in the thermodynamic limit. In this
case, one is generally content to calculate simply the {\em average} value.
This theoretical result should agree with an experiment value, which is done
on a single large sample, within an error that goes to zero in the
thermodynamic limit. 

In this paper, we will be particularly concerned with
spin-spin correlation functions. Clearly, in a random system, one does not have
self averaging of individual correlation functions (though the structure
factor, the sum over all correlations, may be self averaging). Nonetheless,
the tacit assumption is generally made that all reasonable ways of
characterizing the distribution of correlation functions
will give qualitatively similar results, and,
in particular, will give the {\em same} values for critical exponents.
We shall call such behavior of the correlation functions ``conventional''.

We know, however, that there are certain cases where this is not true.
%For example, Derrida\cite{derrida} *** discuss Derrida for 1-d ***.
For example, in certain models for quantum phase transitions with
disorder\cite{quantum}, distributions of correlation functions are so broad
that average and typical values (characterized, say, by the median) behave in
qualitatively different ways both at and away from the critical point. 

In this paper we discuss the question of whether a ``conventional''
description of correlation functions is correct for a classical model
in two dimensions, the random $q$-state Potts model\cite{wu}.
This model has been 
extensively studied by both analytical and numerical
approaches\cite{l-c,cardy,jacobsen,cardy-paris,ludwig,chatelain,chatelain-prl,dotsenko,picco,picco98,jug,chen}.
It is known\cite{baxter}
that the pure model has a second order transition when the number
of Potts states, $q$, is less than or equal to
4 and is first order for $q >4$. The
specific heat exponent of the pure system is positive for $q > 2$, which, from
the Harris criterion\cite{harris}, leads to the conclusion that
disorder will be a relevant perturbation in
this case. Furthermore the transition is {\em always} second order in
two-dimensions\cite{imry-wortis,a-w,h-b}
so disorder has a {\em particularly} strong
effect for $q > 4$,
even changing the order of the transition. The resulting
critical behavior is poorly understood, and its study is one of the main aims
of the present work.

Ludwig\cite{ludwig} has argued that the power law decay of the correlations
{\em at} the critical point is {\em not} governed by a single exponent, $\eta$,
as would be expected if ``conventional'' behavior occurred, but rather shows
``multifractal''\cite{multi} behavior in which there is no simple relation
between the exponents, $\eta_n$, of the  various moments of the correlation
function defined in Eq.~(\ref{moments}) below.
Rather, the whole set of exponents can be conveniently represented
by a function $f(\alpha)$, related to the Legendre transform of $\eta_n$, as we
shall discuss in \S \ref{sec-multi}.  The $\eta_n$ have been calculated
analytically\cite{ludwig} to one-loop order
for $q-2$ small, where the disorder is only weakly
relevant.  Away from the critical point, Ludwig\cite{ludwig} also claims, to
one-loop order in an expansion away from $q=2$,
that
there is only a {\em single} exponent $\nu$, characterizing the divergence of
the correlation length.

Numerical evidence that the correlations at criticality are multifractal has
been provided by the transfer matrix calculations of Jacobsen and
Cardy\cite{jacobsen}.  They showed that cumulants of the log of the
correlation function stay non-zero as the lattice size is increased, which
indicates multifractality. However, their values for $\eta_n$ for $n>0$ are
not obtained directly but are derived from the
cumulant expansion and so become inaccurate for large $n$.
%and they only quote values for $\eta_1$.
In addition, Picco\cite{picco98} has argued that, for the distribution
of disorder used in Ref.~\onlinecite{jacobsen}, the data are in a crossover
region between random and pure behavior, at least for the larger $q$ values,
and so do not give accurate values of the true exponents of the random system.
To our knowledge, the perturbative claim that there is only a single
correlation length exponent $\nu$ has not been verified by other means.

Here we study the $q$-state Potts ferromagnet
by Monte Carlo simulations, using the Wolff
algorithm\cite{wolff} which greatly accelerates equilibration. The main
features of our work are as follows.
\begin{enumerate}
\item
We directly
determine the $\eta_n$ for a wide range of values of $q$ and for several
values of $n$, including $n=0$ (typical), $1$ (average) and others, as well as
the whole distribution of correlation functions.
%Previous work had only presented results for the average correlation function.
We find that $\eta_0$
varies much more strongly with $q$ than does $\eta_1$. Curiously, we find that
within the errorbars $\eta_2$ varies only slightly, if at all, with 
respect to $q$.
\item
We use a different distribution for the disorder from that which is generally
taken. It has the advantage that it is still self dual\cite{k-d},
so the critical
point can be determined exactly, but it is continuous and very broad. As a
result, we argue that it is less susceptible
to crossover effects\cite{picco98,sst} than the 
``bimodal''distributions used before.
\item
We verify explicitly that, at least for $q=8$, the exponents for the
divergence of the average and typical correlation lengths are equal, to within
fairly small numerical errors. Furthermore, to within errors, the value is
the same as that of the pure two-dimensional Ising model. Note, though, that
other aspects of the critical behavior, such as the decay of the correlations
at the critical point, are quite different from those of the pure
two-dimensional Ising model.
% in contrast to the claims of Ref.~\onlinecite{chen}
\end{enumerate}

The layout of the paper is as follows. In \S\ref{sec-model}
we discuss the model and some 
characteristics of the simulations.  In \S\ref{sec-multi} we discuss 
multifractal behavior, which is expected to describe the correlations at the
critical point. Our results {\em at} the critical point are discussed in 
\S\ref{sec-crit} while our results {\em away} from the critical point are
given in
\S\ref{sec-noncrit}. In \S\ref{sec-concl} we summarize our results and give
some perspectives for future work.

\section{The Model}
\label{sec-model}
The Hamiltonian of the $q$-state Potts model is given by
\begin{equation}
\label{ham}
\beta {\cal H} = -\sum_{\langle i, j\rangle} K_{ij} \delta_{n_i n_j} ,
\end{equation}
where each site $i$ on an $N = L\times L$ square lattice is in one of
$q$-states, characterized by an integer $n_i =1,2,\cdots,q$. The couplings,
$K_{ij}$, are positive, and
include the factor of $\beta \equiv 1/k_B T$. They are independent
random variables, drawn from a probability distribution, $P(K)$.

For the pure system, the partition function with coupling $K$ is closely
related to the partition function with the ``dual'' coupling, $K^*$,
where\cite{wu,k-d}
\begin{equation}
( e^K - 1) ( e^{K^*} - 1) = q .
\end{equation}
If $K$ is large (low temperature) then $K^*$ is small (high temperature) and
vice-versa. Assuming that there is a single transition, this
must be at the self-dual point where $K^* = K = K_c$, with
$e^{K_c} = 1 + \sqrt{q}$.

For the random case, the model is still self dual, and hence at its critical
point, provided that the distribution of the dual couplings $P^*$ is
equal to the distribution of the original couplings. One simple example, which
has been extensively used in numerical work, is two delta functions with equal
weight,
\begin{equation}
\label{2delta}
P(K) = {1\over 2} \left[ \delta(K - K_1) + \delta(K - K_2) \right], 
\end{equation}
which is self dual if $K_2 = K_1^*$. Hence Eq.~(\ref{2delta})
describes a family of self dual
distributions characterized by a single parameter, $R \equiv K_1/K_2$.
However, if this ratio is too close to unity, then very large sizes are 
needed\cite{picco98}
otherwise the system is in a ``crossover'' region between the
critical behavior of the pure and the random systems. On the other hand, if the
ratio is made too large, the distribution is fairly close to that of the
percolation problem at criticality, and again the system will be in a crossover
regime.

To avoid these crossovers, it is useful to study the model in Eq.~(\ref{ham})
with a self-dual {\em
continuous} distribution. Although 
Jauslin and Swendsen\cite{j-s} describe how one particular 
continuous self-dual distribution can be constructed, we are not aware of any
results obtained with such a distribution.
Here, we show that with a simple change of variables one can trivially generate
{\em any} self-dual distribution.
In terms of the variable $y$, defined by
\begin{equation}
e^y = {e^K - 1 \over \sqrt{q} },
\end{equation}
the duality condition takes the simple form
\begin{equation}
\label{dual_y}
y^* = -y .
\end{equation}
Hence, expressed in terms of $y$, {\em any} even distribution,
$P_Y(y)$, is self-dual.
Note that because Eq.~(\ref{dual_y}) is so
simple the Jacobian in the transformation from the distribution of $y$
to the distribution of $y^*$ is unity, unlike the situation going from the
distribution of $K$ to that of $K^*$, where the Jacobian is non-trivial.

\begin{figure}
\epsfxsize=\columnwidth\epsfbox{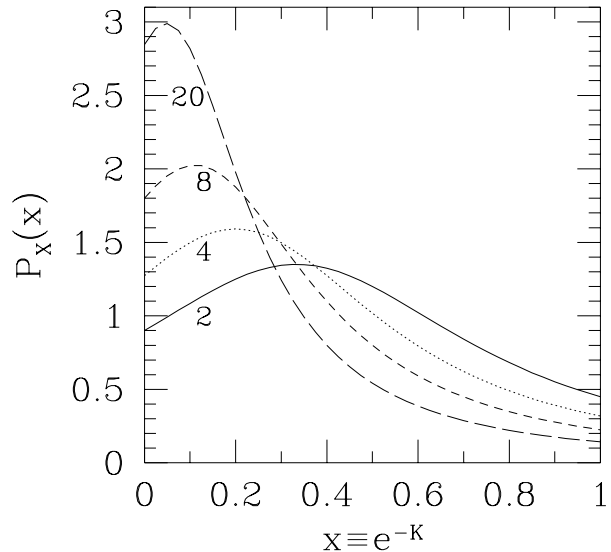}
\caption{
A plot of the self-dual distribution used in the simulations at the critical
point, for several values of $q$.
It is given by Eq.~(\protect\ref{px}) of the text.
}
\label{dist}
\end{figure}

The quantity that enters in statistical mechanics averages is $x \equiv
e^{-K}$, which, for the ferromagnetic couplings discussed here,
takes values in the range from 0 to 1. In order that the
model has strong disorder we seek a distribution of $x$ that is non-zero
everywhere, and has a finite weight at the end points. This means that
$P_Y(y) \propto \exp(-|y|)$
for $y \to \pm \infty$. A convenient choice, which we
use in the rest of this paper for simulations at criticality, is
\begin{equation}
P_Y(y) = {1\over \pi}\, {\rm sech}(y) ,
\end{equation}
for which the corresponding distribution for $x \equiv e^{-K}$ is
\begin{equation}
P_X(x) =  {2 \over \pi} { \sqrt{q} \over (1-x)^2 + q x^2 } .
\label{px}
\end{equation}
This is plotted in Fig.~\ref{dist} for several values of $q$. Another advantage
of this distribution is that it is easy to
generate random numbers with probability $P_X(x)$ (which are needed for the
Wolff algorithm\cite{wolff}). If $r$ is a random number
with a uniform distribution between 0 and 1, one simply takes $x$ to be
\begin{equation}
x = {1 \over 1 + \sqrt{q} \tan (\pi r / 2) } .
\end{equation}

We apply periodic boundary conditions. The numbers of samples used are shown in Table~\ref{no-of-samples}, and the number of Monte Carlo
sweeps are shown in Table~\ref{no-of-sweeps}. Note that the number of sweeps for
averaging is quite large, given that we use the Wolff algorithm which
equilibrates the system quickly even for large sizes. The reason is that we
wish to obtain the whole distribution of spin-spin correlation functions, 
so {\em each} pair correlation function has to be obtained with high
precision. We cannot use
noisy data for the individual correlation functions and
rely on averaging over a large number of pairs to improve
the statistics. This would be fine for the average but lead to {\em systematic}
errors for other moments.  For each sample we determine the correlation
function for spins $L/2$ apart
for a large number of pairs, depending on $L$, as indicated
in Table~\ref{no-of-pairs}.

%\newpage
\begin{table}
\begin{tabular}{|c|c|c|c|c|c|c|}
q & L=16 & L=32 & L=64 & L=128 & L=256 & L=512\\
\hline
2 & 100 & 100 & 200 & 125 & 18 & -\\ 
3 & 3000 & 1200 & 400 & 400 & 365 & 95 \\
%\hline
4 & 2000 & 600 & 210 & 200 & 125 & 50 \\
%\hline
5 & 2000 & 600 & 200 & 200 & 100 & - \\
%\hline
8 & 1500 & 1000 & 904 & 977 & 426 & 87 \\
%\hline
20 & 1000 & 600 & 513 & 500 & 265 & 67 \\
\end{tabular}
\caption
{
The number of samples used for each value of $q$ and $L$.
\label{no-of-samples}
}
\end{table}

\begin{table}
\begin{tabular}{|c||c|c|} 
q & $L \le 256$ & L=512\\
\hline
$2 - 8$ & 50000(500)& 100000(1000) \\
20 & 250000(2500) & 250000(2500) \\
\end{tabular}
\caption{
The number of Monte Carlo sweeps for averaging.
In brackets is the number of sweeps
for equilibration.
Note that for $q = 8$, $L = 512$,  43 of the 87 samples
were done with 50000(500) sweeps.
\label{no-of-sweeps}
}
\end{table}

\begin{table}
\begin{tabular}{|c|c|c|c|c|c|}
L=16 & L=32 & L=64 & L=128 & L=256 & L=512\\
\hline
256 & 1024 & 4096 & 1638 & 655 & 10000 \\
\end{tabular}
\caption
{
The number of pairs used in the calculation of $C(L/2)$ for different values of
$L$.
\label{no-of-pairs}
}
\end{table}

%\newpage

%\begin{center}
\section{Multifractal Behavior}
\label{sec-multi}
Consider the spin-spin correlation function, $C_{i,j}$,
%$C(r)$,
for a pair of sites,
$i$ and $j$. If the spins are 
separated by a distance $r$, we shall also denote this by $C(r)$, where
\begin{equation}
C(r) = {q \over q-1} \left\langle  \delta_{n_i n_j} - {1\over q} \right\rangle ,
\end{equation}
in which $\langle \cdots \rangle$ denotes a thermal average for a single
realization of the disorder. 
At the critical point, the correlations decay with a power law and we find it
convenient to define a set of exponents $\eta_n$ by
\begin{equation}
[ C(r)^n]^{1/n}_{\rm av}  \sim r^{-\eta_n}  ,
\label{moments}
\end{equation}
for $n = 0, 1, 2, \cdots, $ where
averages over disorder are indicated by
$[ \cdots ]_{\rm av}$. The $n=0$ value in Eq.~(\ref{moments}),
which gives the behavior of a
``typical'' correlation function, is obtained as the limit
$n\to 0$, {\em i.e.}\/ $\exp[\ln C(r)]_{\rm av}$.
A ``typical'' correlation function can also be defined as 
the median of the distribution, with very similar results. 

In what we are calling ``conventional behavior'' all the $\eta_n$ are
equal. However,
according to Ludwig\cite{ludwig}, there is no simple relation between the
$\eta_n$ for the random Potts model in two dimensions. Instead one has
multifractal behavior in which 
the probability distribution of the $C(r)$ is given by
\begin{equation}
\tilde{P}(\alpha) = {\cal N} \exp[- f(\alpha) \ln r]
\label{falpha-eq}
\end{equation}
where
\begin{equation}
\alpha =  -{\ln C(r) \over \ln r} ,
\end{equation}
and $\cal N$ is the normalization.
From general considerations, $f(\alpha)$ must have a minimum at some point,
$\alpha_0$ say, and the value of $f(\alpha_0)$
can be absorbed into the normalization, so we set $f(\alpha_0) = 0$.

In the thermodynamic limit, averages can be done by a saddle point method,
with the result that for each $n$, there is a corresponding value of $\alpha$
given by
\begin{equation}
f^\prime(\alpha) = -n
\label{fprime}
\end{equation}
and then
\begin{equation}
n \eta_n = f(\alpha) + \alpha n .
\label{legendre}
\end{equation}
The error in the saddle point calculation is of order\cite{ludwig}
$1/\sqrt{\ln r}$ which
falls off extremely slowly with distance. This will be important in the
analysis of the numerical results in the next section.

%\newpage
\begin{figure}
\epsfxsize=\columnwidth\epsfbox{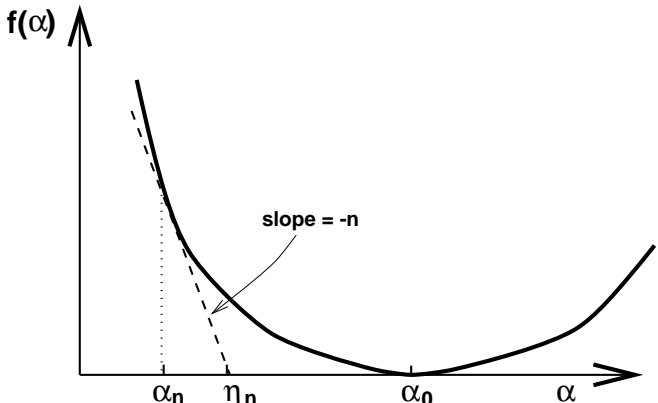}
\caption{
A sketch of the function $f(\alpha)$ which characterizes the multifractal
nature of the distribution of the correlation functions. To determine the
exponent $\eta_n$, describing the power law decay of the $n$-th moment of the
correlations at the critical point, according to Eq.~(\protect\ref{moments}),
locate
the point where the slope of the curve is $-n$ and draw the tangent at that
point. Where the tangent intersects the horizontal axis is $\eta_n$. Clearly,
$\eta_0 = \alpha_0$, the location of the minimum.
}
\label{falpha}
\end{figure}

Eqs.~(\ref{fprime}) and (\ref{legendre}) imply 
a simple graphical relation between $f(\alpha)$ and the $\eta_n$,
illustrated in Fig.~\ref{falpha}. One locates the point on the $f(\alpha)$
curve with slope $-n$, and where this intersects the horizontal axis is
$\eta_n$. Clearly then $\eta_0 = \alpha_0$, the location of the minimum.
Because the $\eta_n$ cannot be negative (otherwise the correlations would grow
with distance), the function $f(\alpha)$ will diverge as $\alpha$ approaches
some non-negative value.
%Positive moments do not give information on the
%values of $f(\alpha)$ for $\alpha > \alpha_0$, which are related to the weight
%of the tail of the distribution of the $C(r)$ for small $C$.

\section{Results at Criticality}
\label{sec-crit}
\begin{figure}
\epsfxsize=\columnwidth\epsfbox{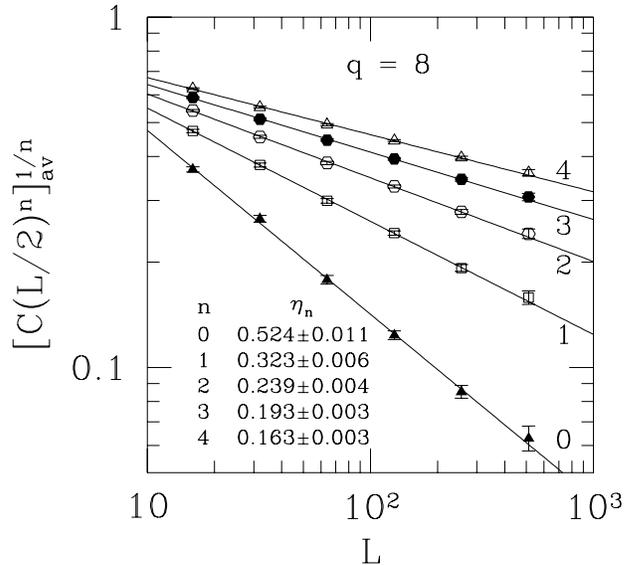}
\caption{
Data for the $q=8$ model at criticality for sizes between
$L=16$ and 512, for $n=0,1,2,3$ and $4$.
The lines are fits to the data and the slopes give the values for
$\eta_n$ indicated.
}
\label{eta8}
\end{figure}

We concentrate on the correlation function between two spins as far apart as
possible in the lattice in either the horizontal or vertical direction,
{\em e.g.}\/ if one site is at (0,0) the other is at
$(0,L/2)$ or $(L/2,0)$. Fig.~\ref{eta8} shows our results 
for $[C(L/2)^n]_{\rm av}^{1/n}$ for several
values of $n$ at criticality for the case of $q=8$.  The slopes are equal to
$-\eta_n$. In the ``conventional'' picture, they would all be equal. Clearly
this is not the case; rather the slopes change with $n$ as expected for
multifractal behavior.

%\newpage

\begin{figure}
\epsfxsize=\columnwidth\epsfbox{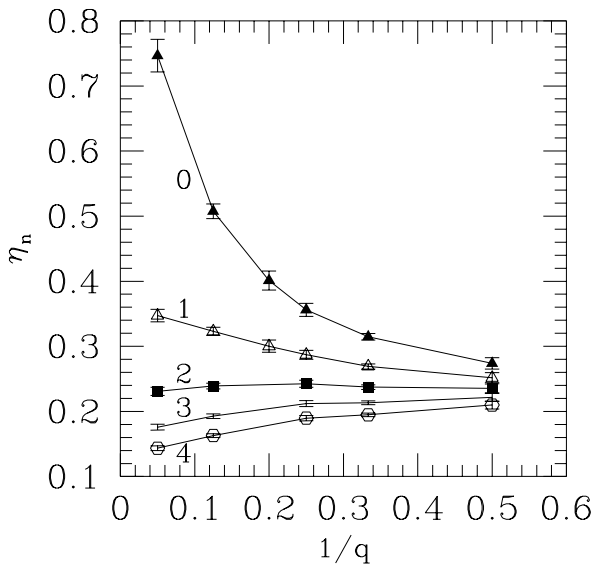}
\caption{
Results for $\eta_n$, for several values of $n$, plotted against $1/q$.
}
\label{etan_q}
\end{figure}

Fig.~\ref{etan_q} summarizes our results for the exponents $\eta_n$, which are
also presented in numerical form in Table~\ref{eta_results}. Notice, from
Fig.~\ref{eta8}, that $[C^n]_{\rm av}^{1/n}$ is an increasing function of $n$,
as expected on general grounds\cite{ludwig} since
$\ln [C^n]_{\rm av}$ is a convex function\cite{feller} of $n$. This 
implies that, for fixed $q$, $\eta_n$ decreases with increasing $n$, as seen
in Fig.~\ref{etan_q} and Table~\ref{eta_results}.
Clearly $\eta_0$, representing the decay of typical correlation
functions, varies very strongly with $q$, while the other $\eta_n$
vary less strongly.
For small $n$, $\eta_n$ increases as $q$ increases, whereas for large $n$ the
converse is true. Within errors, $\eta_2$ is independent of $q$ for the range
studied. 
%Notice that for fixed $q$, $\eta_n$ decreases with increasing
%$n$, as expected on general grounds\cite{ludwig}, since
%$\ln [C^n]_{\rm av}$ is a convex function\cite{feller} of $n$ and so
%$[C^n]_{\rm av}^{1/n}$ is an increasing function of $n$.

\begin{table}
\begin{tabular}{|l|r|l||l|r|l||l|r|l|}
%\hline\hline
n  & q &     $\eta_n$ & n & q &  $\eta_n$ & n & q & $\eta_n$\\
\hline\hline
0 & 2  & 0.274(9)  & 1 & 2  & 0.252(8) & 2 & 2  & 0.235(7)  \\
0 & 3  & 0.315(5)  & 1 & 3  & 0.269(4) & 2 & 3  & 0.237(4)  \\
0 & 4  & 0.356(10)  & 1 & 4  & 0.287(7) & 2 & 4  &  0.242(5) \\
%0& 5  & 0.401(14) & 1 & 5  & 0.300(9 \\
0 & 8  & 0.507(11) & 1 & 8  & 0.323(6) & 2 & 8  &  0.239(4) \\
0 & 20 & 0.746(25) & 1 & 20 & 0.347(10)& 2 & 20 &  0.231(6) \\
\hline
3 & 2  & 0.222(6)  & 4 & 2  & 0.210(6) &   &    &           \\
3 & 3  & 0.213(3)  & 4 & 3  & 0.195(3) &   &    &           \\
3 & 4  & 0.212(5)  & 4 & 4  & 0.190(4) &   &    &           \\
3 & 8  & 0.193(3)  & 4 & 8  & 0.163(3) &   &    &           \\
3 & 20 & 0.176(5)  & 4 & 20 & 0.144(4) &   &    &           \\
%\hline\hline
\end{tabular}
\caption
{
\label{eta_results}
The values of $\eta_n$, defined by Eq.~(\protect\ref{moments}) at criticality,
for different values of $q$ and $n$. $n=1$ corresponds
to the average correlation function and $n=0$ to the typical correlation
function. The values for $n=0$ are obtained by averaging results for the
exponential of the average of the log and the median. These agreed within
expected statistical fluctuations. 
The quantity in brackets is the error in the last decimal place.
}
\end{table}

%\newpage

It would be of interest to
investigate the limit $q \to \infty$, but we are not able to equilibrate $q$
values very much larger than 20. However, a naive extrapolation of our data
gives $\lim_{q\to\infty} \eta_1 = 0.37 \pm 0.01$.

Note that the data in Fig.~\ref{etan_q} for $q=2$ show a small but non-zero
variation of $\eta_n$ with $n$. However, $q=2$ is expected to have
marginal behavior, in which $\eta_n = 1/4$ (pure value) for all
$n$ but with logarithmic corrections\cite{ludwig},
{\em i.e.}\/  $[C(r)^n]_{\rm av}^{1/n} \sim r^{-1/4} \ln(r/r_0)^{\lambda_n}$,
where, for example, $\lambda_0 = -1/8$ and $\lambda_1= 0$.
The data cannot unambiguously
determine the form of the logarithmic factors (though it is {\em consistent}
with them for appropriate values of $\lambda_n$ and $r_0$). 
The fits used to get the data in Fig.~\ref{etan_q} therefore
assume a pure power law
behavior, and the apparent small change of $\eta_n$ with $n$ for $q=2$ is
presumably due to neglect of the logarithmic corrections. Note in
Table~\ref{eta_results}, that for $n=1, q=2$, (where the logarithmic
corrections are predicted to be absent\cite{ludwig})
we get $\eta_1$ very close to the
pure Ising value of 1/4, as expected.

%\newpage
\begin{figure}
\epsfxsize=\columnwidth\epsfbox{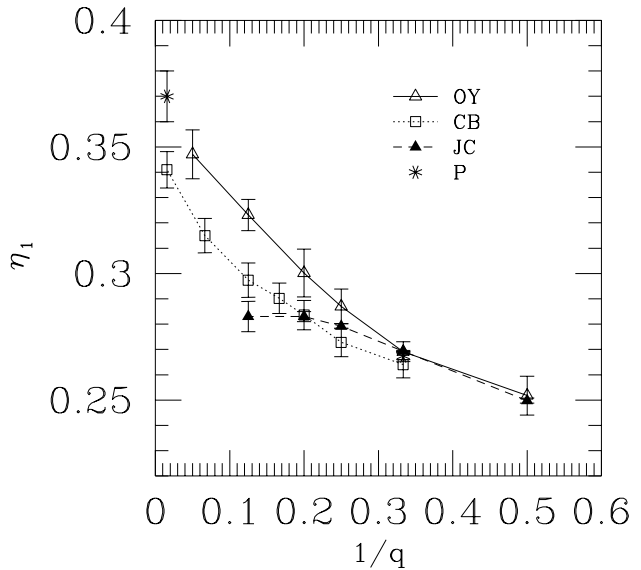}
\caption{
A comparison between our (OY) estimates of $\eta_1$ (relevant for the average
correlation function), with those of Jacobsen and Cardy\protect\cite{jacobsen}
(JC), Monte Carlo results of
Chatelain and Berche\protect\cite{chatelain} (CB) and
Picco\protect\cite{picco,picco98} (P).
Ref.~\protect\onlinecite{chatelain} also has
data from Transfer Matrix techniques which agree with their Monte Carlo data
within the errors. 
}
\label{other_eta1}
\end{figure}

In Fig.~\ref{other_eta1} we compare our results for $\eta_1$ with those of
other authors. An extrapolation of results to larger
values of $q$ is consistent with 
the result of Picco\cite{picco98} for $q=64$. Our results also agree with
Picco's\cite{picco} for $q=3$.
For larger $q$ values they lie above those of
Jacobsen and Cardy\cite{jacobsen}. This is not surprising in view of
Picco's\cite{picco98} claim that the strength of disorder used by Jacobsen and
Cardy is not large enough for $q=8$ to be in the asymptotic critical regime. 
The bimodal distribution used in those studies is characterized by the ratio 
$R$ of the two interactions. Ref.~\onlinecite{jacobsen} used $R=2$ whereas
Picco argues that $R \simeq 10$ is needed.
More surprisingly, our results also lie somewhat
above those of Chatelain and Berche\cite{chatelain} who used stronger
disorder, $R\ge 10$ for $q \ge 8$. Perhaps in that case there is a
%For such a large value of $R$, perhaps there is
crossover from the percolation critical point. 
%Perhaps there are still some residual crossover effects for
%the bimodal distribution.
Our results do, however, also agree with the
result (not shown on the figure) of Wiseman and Domany\cite{w-d} that, for
$q=4$,
$\eta_1 = 0.290 \pm 0.006$. For $q=3$, our results
also agree with the analytical three-loop
calculation of Dotsenko et al.\cite{dotsenko}, though the difference between
the random and pure value is not very great in this case.

%We are not aware of any other {\em quoted} values for $\eta_n$ with $n \ne 1$. 
%However, Jacobsen and Cardy\cite{jacobsen}
%present results from which $\eta_0$ can be
%{\em inferred}\/\cite{jc_table} for $q=3$
%and $8$.
%However, we find that any reasonable extrapolation of their data
%to infinite system size
%gives results which are significantly below ours, especially for $q=8$. This
%may be related to Picco's remarks that the ratio of interactions, $R=2$, used
%in Ref.~\onlinecite{jacobsen} is too small, at least for $q=8$.

There are few studies of any other values for $\eta_n$ with $n \ne 1$.
Conformal field theory has been employed by Lewis \cite{lewis} to obtain
second order expansions for $\eta_n$; for $q = 3$ he calculates 
$\eta_0 = 0.314$, $\eta_2 = 0.236$ (which agrees with a similar work by 
Dotsenko et al.\cite{ddp}), and $\eta_3 = 0.220$. These values agree quite
well with ours.
Jacobsen and Cardy\cite{jacobsen} present results from which $\eta_0$ can be
{\em inferred}\/\cite{jc_table} for $q=3$ and $8$.
However, we find that any reasonable extrapolation of their data
to infinite system size
gives results which are significantly below ours, especially for $q=8$. This
may be related to Picco's remarks\cite{picco98}
that the ratio of interactions, $R=2$, used
in Ref.~\onlinecite{jacobsen} is too small, at least for $q=8$. 
Dotsenko et al.\cite{ddp} perform Monte Carlo simulations 
estimating $\eta_2 = 0.227(2)$ for both $q=3$ and $q=4$.
A more detailed 
analysis\cite{picco_pc} of this data yields
$0.23(2)$ for $q=3$ and $0.229(2)$ for 
$q=4$.  Although these values
are slightly smaller than ours, they support the trend that $\eta_2$ is 
barely changing with $q$.

The exponent characterizing the relevance of weak disorder is
$y \equiv \alpha_P / \nu_P$, where $\alpha_P$ and $\nu_P$ are
the specific heat and correlation length exponents of the pure system. This
vanishes as $q \to 2^+$ and has the value $y=2/5$ for $q=3$.
Ludwig\cite{ludwig} has shown that for $(n-1)y \ll 1$ 
\begin{equation}
\eta_n - \eta_1 \simeq -{y\over 8} (n-1) .
\end{equation}
For $q=3$ this gives $\eta_0 - \eta_1 = 0.05$, which agrees well with our
result
of $0.046$ (with an uncertainty of around 0.006), see Table \ref{eta_results}.
This is in contrast to Jacobsen and Cardy\cite{jacobsen} who stated that their
data did not agree with this relation, though they did not quote a value for
$\eta_0$.

\begin{figure}
\epsfxsize=\columnwidth\epsfbox{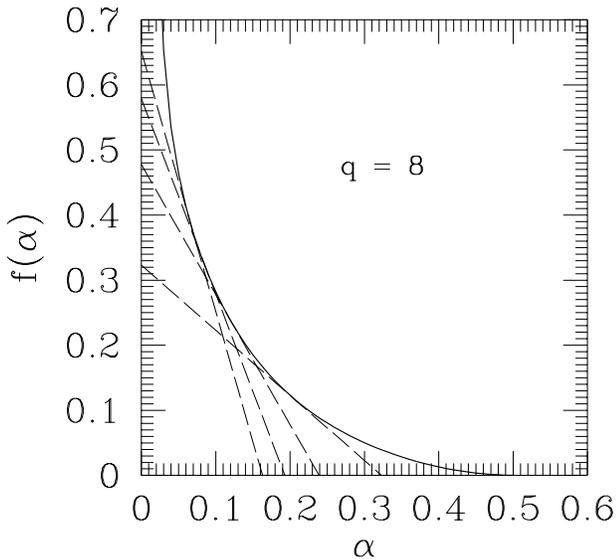}
\caption{
The solid line is a
reconstruction, for $q=8$,
of the multiscaling function, $f(\alpha)$. It is obtained as
the tangent to the dashed lines, whose formulae are $n(\eta_n-\alpha)$,
for $n=0,1,2,3$ and $4$, see
Eqs.~(\protect\ref{fprime})
and (\protect\ref{legendre}). See also Fig.~\protect\ref{falpha}.
}
\label{falpha_q8}
\end{figure}

From the values of the exponents $\eta_n$, we
can construct the multiscaling function $f(\alpha)$
from.~(\ref{fprime}) and (\ref{legendre}).
This is shown in Fig.~\ref{falpha_q8} for $q=8$.

%\newpage

%\newpage
\begin{figure}
\epsfxsize=\columnwidth\epsfbox{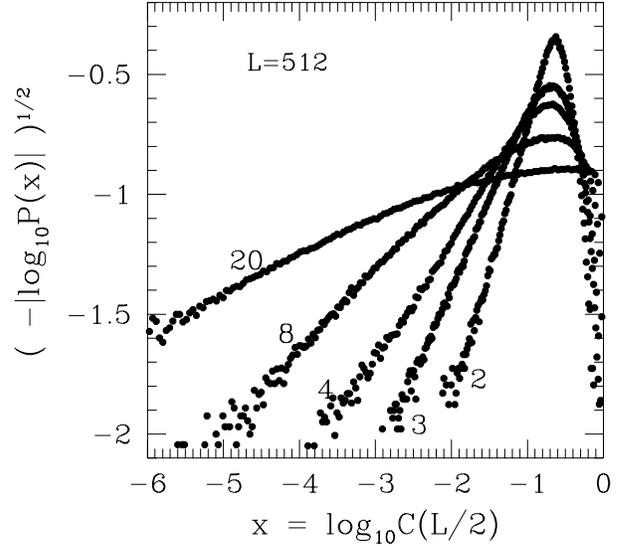}
\caption{
The distribution of
$\log_{10} C(L/2)$, for $L=512$ and several different values of $q$ at
criticality.
}
\label{plogc2_L512}
\end{figure}

So far, we have discussed results for the moments of the correlation
function. 
Now, in Fig.~\ref{plogc2_L512}, we show results for the whole 
{\em distribution} of 
$\log_{10} C(L/2)$, for $L=512$ and several different values of $q$ at
criticality. The vertical scale is chosen so that, for a log-normal 
distribution, the tails of the curves would be two straight lines symmetric
about the peak. The data for $q=2$ is not very different from this, but for
large $q$, especially $q=20$, the curve is not only much broader but also
very asymmetric. Furthermore, even in the
tail to the left of peak, the data for $q=20$ is significantly curved.
It would be interesting to understand the behavior
of the distribution in the limit $q\to \infty$. 

\begin{figure}
\epsfxsize=\columnwidth\epsfbox{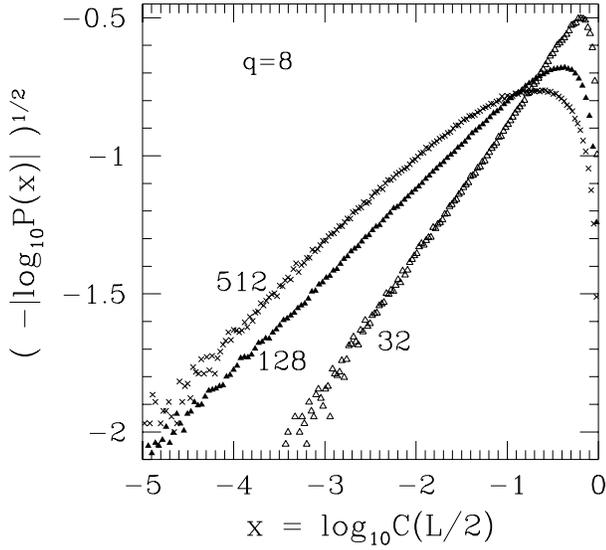}
\caption{
The distribution of
$\log_{10} C(L/2)$, for $q=8$ and several different values of $L$ at
criticality.
}
\label{plogc_q8}
\end{figure}

Similar data is shown in Fig.~\ref{plogc_q8} for $q=8$ and different values of
$L$ at criticality. As expected, the distribution becomes broader for
increasing $L$. For small $L$ the data for small $C(L/2)$ is roughly straight
indicating that this tail follows close to a log-normal distribution. However, 
for $L=512$, significant curvature in the tail
can be seen. Furthermore, attempts to fit to
Eq.~(\ref{falpha-eq})
to directly determine the multiscaling function $f(\alpha)$ (and hence
to compare
with the reconstruction of this function from the moments in
Fig.~\ref{falpha_q8})
were unsuccessful.
Eqs.~(\ref{fprime}) and (\ref{legendre})
and Fig.~\ref{falpha_q8} depend upon a saddle point
approximation which is only valid for $\sqrt{\ln (L/2)} \gg 1 $, which is not
realizable in a Monte Carlo simulation. 

%\newpage

%\newpage
\begin{figure}
\epsfxsize=\columnwidth\epsfbox{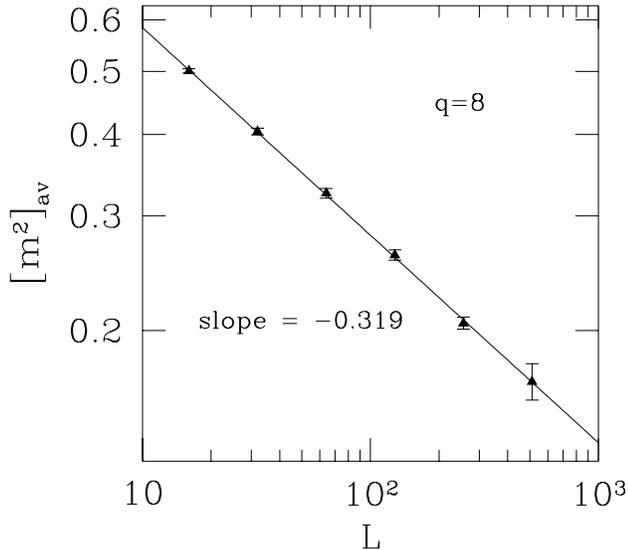}
\caption{
Data for $[m^2]_{\rm av}$, defined by Eq.~(\protect\ref{m2-def}),
at the critical point for $q=8$. From the slope,
which is $-\eta_1$, see Eq.~(\ref{m2-fss}), we get $\eta_1 = 0.319 \pm 0.007$.
}
\label{m2}
\end{figure}

As a check on our results we also determined $\eta_1$ from the mean square
magnetization at the critical point. In $d=2$, finite size scaling predicts
\begin{equation}
\label{m2-fss}
[ \langle m^2 \rangle ]_{av} \sim L^{-2 \beta/\nu} = L^{-\eta_1}
\end{equation}
where $\beta$ is the order parameter exponent, $\nu$ the correlation length
exponent, 
\begin{equation}
\label{m2-def}
m^2 = {1 \over N^2} \sum_{i,j} C_{i,j} 
 = {q \over q-1} \left\langle \sum_{n=1}^q \rho_{n}^2 - {1\over q}
\right\rangle,
\end{equation}
and $\rho_{n}$ is the fraction of sites in state $n$.

\begin{table}
\begin{tabular}{r l c}
\ \ \ \ \ \ \ \ \ \ \ \ \ \ \ \ \ \ q  &     $\eta_1$ &\hfill \\
\hline
\ \ \ \ \ \ \ \ \ \ \ \ \ \ \ \ \ \ 3  & 0.259(4) &\hfill \\
\ \ \ \ \ \ \ \ \ \ \ \ \ \ \ \ \ \ 8  & 0.319(7) &\hfill\\
\ \ \ \ \ \ \ \ \ \ \ \ \ \ \ \ \ \ 20 & 0.345(9) &\hfill\\
\end{tabular}
\caption
{
\label{eta-m2-tab}
The values of $\eta_1$, obtained from the mean square magnetization,
Eq.~(\protect\ref{m2-fss}), at criticality,
for different values of $q$.
}
\end{table}

For $q=8$ the data is shown in Fig.~\ref{m2}
and the values of $\eta_1$ determined
this way are given in Table~\ref{eta-m2-tab}.
The results agree with those found from the
correlation function $C(L/2)$, see Table~\ref{eta_results},
to within statistical errors.

\section{Results away from Criticality}
\label{sec-noncrit}

Although the distribution of correlation functions at the critical point
has multifractal behavior, Ludwig\cite{ludwig}
claims that only a single exponent $\nu$
describes the divergence of the correlation length as the critical point is
approached.

%Although several
%authors\cite{jacobsen,chen,picco,chatelain,chatelain-prl}
%have {\em numerically} determined the 
%exponent for the divergence of the
%correlation length of the average correlation function, there are,
%to our knowledge, no numerical determinations of the correlation length
%for other moments, such as $n=0$ (the typical correlation
%function).

%\newpage
\begin{figure}
\epsfxsize=\columnwidth\epsfbox{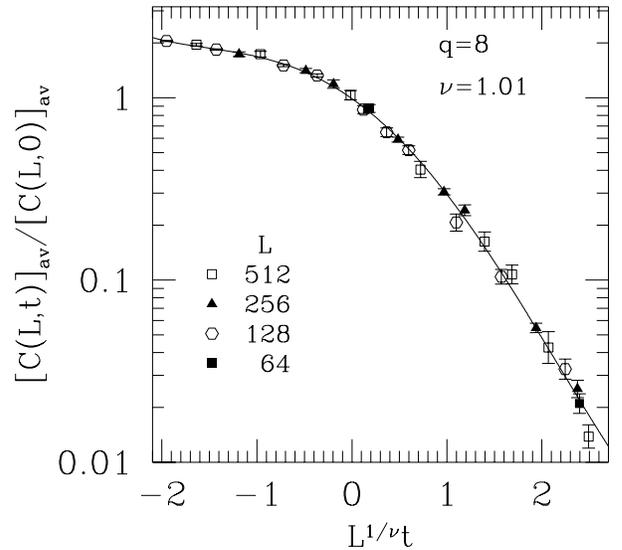}
\caption{
A scaling plot of the data for the average correlation function away from the
critical point for $q=8$. Here $t$, defined by Eq.~(\protect\ref{t}), is the
deviation from criticality. The solid curve is a polynomial fit.
}
\label{nu_mean}
\end{figure}

%\newpage
\begin{figure}
\epsfxsize=\columnwidth\epsfbox{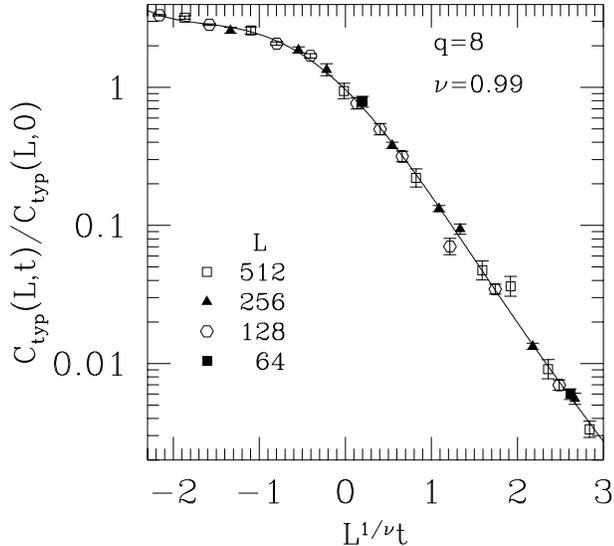}
\caption{
A scaling plot of the data for the typical correlation function away from the
critical point for $q=8$. Here $t$, defined by Eq.~(\protect\ref{t}), is the 
deviation from criticality. The solid curve is a polynomial fit.
}
\label{nu_typical}
\end{figure}

There are {\em analytical} predictions\cite{l-c,dotsenko,jug}
for how $\nu$ varies with
$q$ for $q$ near 2. The deviation of these results from the value for
the pure Ising model, $\nu=1$, is
small, and numerical studies have so far been unable to resolve it.
Here we concentrate on $q=8$, where disorder changes the order of
the transition from first to second and so the perturbative analytical
approach is not applicable.

We define 
\begin{equation}
t = (T-T_c)/T_c , 
\label{t}
\end{equation}
so the Boltzmann factors, $x \equiv
e^{-K}$, are first calculated at the critical point from the self-dual
distribution in Eq.~(\ref{px}), and are then
modified by
the replacement $x \to x^{1/(1+t)}$ away from criticality.

According to finite size scaling, the average correlation function away from
criticality, $[C(L/2, t)]_{\rm av}$, is related to that at criticality by
\begin{equation}
{ [C(L/2, t)]_{\rm av} \over [C(L/2, 0)]_{\rm av} } = f(\L^{1/\nu} t) ,
\end{equation}
where $f$ is a scaling function.
There is an analogous expression for the typical correlation function.
Figs.~\ref{nu_mean} and \ref{nu_typical} show scaling plots,
assuming this form, in which $\nu$ has been adjusted to get the best data
collapse.
From the chi-squared of the fits we estimate
\begin{eqnarray}
\nu & = & 1.01   \pm 0.02 \qquad \mbox{average} \\
\nu & = & 0.99   \pm 0.02 \qquad \mbox{typical} ,
\end{eqnarray}
showing that the values of $\nu$ agree to within the errors. 

%Several earlier studies\cite{jacobsen,chen,chatelain-prl}
%had also found that, for the {\em average} correlation function, $\nu$ is
%close to 1, the value for the pure two-dimensional Ising model.
We believe that this is the first numerical calculation which verifies
that the exponents for the average and typical correlation length are equal.
This indicates that the correlations have conventional, rather than
multifractal, behavior away from the critical point.
It is also interesting that $\nu$ is close to (and perhaps equal to) 1,
the value for the pure two-dimensional Ising model.

\section{Conclusions}
\label{sec-concl}

In this work we have presented results for the critical exponents of the
random $q$-state Potts model, for various moments of the correlation
functions, at and away from the critical point. We have confirmed in greater
detail than before that that the correlations have multifractal behavior at the
critical point. We have also verified, for the first time, that there is only a
single exponent $\nu$ describing the divergence of the correlation length.

Implicit in our discussion has been the assumption of universality {\em i.e.}\/
that all the exponents $\eta_n$, and hence the multifractal function,
$f(\alpha)$, do not depend on details of the model, such as the form of the
distribution of disorder. While universality is well established for pure
systems, to our knowledge, it has not been convincingly demonstrated in
situations where there are multifractal correlations.

Another issue which merits further discussion is the form of corrections
to scaling. Generally these have a power law form, in which the 
exponent is that of the leading irrelevant operator. However, when there are
multifractal correlations with a continuous spectrum of exponents
characterized by $f(\alpha)$, is it possible that the approach to the
asymptotic limit is slower? This is certainly the
case for $f(\alpha)$ itself, where
corrections fall off only as $1/\sqrt{\ln r}$, much slower than a power law.
Whether the same slow decay of corrections also applies to the exponents
$\eta_n$ is not clear to us.

We note that in spin glasses, the numerical
results, particularly for $\eta$, do not seem to satisfy
universality\cite{campbell}. Could a resolution be that the approach to the
thermodynamic limit is only logarithmic, so astronomically large sizes 
are needed to see universality?

\acknowledgments
We would like to thank J.~Deutsch for useful discussions on multifractals,
and J.~Cardy for helpful comments on the manuscript.
This work was supported by the National Science Foundation under grant DMR
9713977.
%Mention the CLC ?

%\end{center}


\begin{references}

%\bibitem{derrida}
%    B.~Derrida, Phys. Rep. {\bf 103}, 29 (1984);  B.~Derrida and H.~Hilhorst,
%    J. Phys. C. {\bf 14} L539 (1981).

\bibitem{quantum}
    D.~S.~Fisher, Phys. Rev. Lett. {\bf 69}, 534 (1992); Phys. Rev. B {\bf
    51}, 6411 (1995); C.~Pich A.~P.~Young, H.~Rieger and N.~Kawashima, Phys.
    Rev. Lett, {\bf 81}, 5916, (1998).

\bibitem{wu}
    For a review of the Potts model, see F.~Y.~Wu, Rev. Mod. Phys. {\bf 54},
    235 (1982).

\bibitem{l-c}
    A.~W.~W.~Ludwig and J.~Cardy,  Nucl. Phys. B, {\bf 285}, 687 (1987).

\bibitem{cardy}
    J.~Cardy and J.~J.~Jacobsen, Phys. Rev. Lett. {\bf 79}, 4063 (1997).

\bibitem{jacobsen}
    J.~J.~Jacobsen and J.~Cardy, Nucl. Phys. B, {\bf 515}, 701 (1998).

\bibitem{cardy-paris}
    J.~Cardy, cond-mat/9806355.

\bibitem{ludwig}
    A.~W.~W.~Ludwig, Nucl. Phys. B, {\bf 330}, 639 (1990).

\bibitem{chatelain}
    C.~Chatelain and B.~Berche, cond-mat/9902212.

\bibitem{chatelain-prl}
    C.~Chatelain and B.~Berche, Phys. Rev. Lett. {\bf 80}, 1670 (1998).

\bibitem{dotsenko}
    V.~Dotsenko, M.~Picco and P.~Pujol, Nucl. Phys. B, {\bf 455}, 701 (1995).

\bibitem{picco}
    M.~Picco, Phys. B, {\bf 54}, 14930 (1996); Phys. Rev. Lett. {\bf 79} 2998
    (1997).

\bibitem{picco98}
    M.~Picco, cond-mat/9802092

\bibitem{jug}
    G.~Jug and B.~N.~Shalaev, Phys. Rev. B, {\bf 54}, 342 (1996).

\bibitem{chen}
    S.~Chen, A.~M.~Ferrenberg and D.~P.~Landau, Phys. Rev. E, {\bf 52}, 1377
    (1995). 
    
\bibitem{baxter}
    R.~J.~Baxter, J. Phys. C, {\bf 6} L445 (1973).

\bibitem{harris}
    A.~B.~Harris, J. Phys. C {\bf 7}, 1671 (1974).

\bibitem{imry-wortis}
    Y.~Imry and M.~Wortis, Phys. Rev. B {\bf 19}, 3580 (1979).

\bibitem{a-w}
    M.~Aizenman and J.~Wehr, Phys. Rev. Lett. {\bf 62}, 2503 (1989).

\bibitem{h-b}
    K.~Hui and A.~N.~Berker, Phys. Rev. Lett. {\bf 62}, 2507 (1989).

\bibitem{multi}
    T.~C.~Halsey, M.~H.~Jensen, L.~P.~Kadanoff, I.~Procaccia, and
    B.~I.~Shraiman, Phys. Rev. A, {\bf 33}, 1141 (1986), and references
    therein.

\bibitem{wolff}
    U.~Wolff, Phys. Rev. Lett. {\bf 62}, 361 (1989); R.~H.~Swendsen and
    J.~Wang, Phys. Rev. Lett. {\bf 58}, 86 (1987).

\bibitem{k-d}
    W.~Kinzel and E.~Domany, Phys. Rev. B {\bf 23} 3421 (1981).

\bibitem{sst}
    W.~Selke, L.~N.~Shchur, and A.~L.~Talapov, in {\em Annual Reviews of
    Computational Physics}, Vol. 1, Ed. D. Stauffer (World Scientific), p.17.

\bibitem{j-s}
    H.~R.~Jauslin and R.~H.~Swendsen, Phys. Rev. B {\bf 24}, 313 (1981).

\bibitem{feller}
    W.~Feller, An Introduction to Probability Theory and its Applications, 
    Vol. 2, p. 153, J Wiley and Sons, New York (1966).

\bibitem{w-d}
    S.~Wiseman and E.~Domany, Phys. Rev. E {\bf 51}, 3074 (1995).

%added the next three - terry
\bibitem{lewis}
    M.~Lewis, Europhys. Lett. {\bf 43}, 189 (1998); (private communication).

\bibitem{ddp}
    V.~Dotsenko, V.~Dotsenko and M.~Picco, Nucl. Phys. B {\bf 520}, 663 (1998). 

\bibitem{jc_table}
    In Table~6 of Ref.~\protect\onlinecite{jacobsen}, the column marked ``1.
    cumulant'' gives a value for $\eta_0$ for a strip of width $L$, when
    multiplied by $L^2/\pi$. This should then be extrapolated to $L=\infty$.

\bibitem{picco_pc}
    M.~Picco, private communication.

\bibitem{campbell}
    L.~W.~Bernadi, S.~Prakash and I.~A.~Campbell, Phys. Rev. Lett. {\bf 77},
    2798 (1996); L.~W.~Bernadi and I.~A.~Campbell, Phys. Rev. B {\bf 56}, 5271
    (1997).  


    
\end{references}
\end{document}